\documentstyle[12pt]{article}
\newcommand{\ka}{\kappa}
\newcommand{\la}{\lambda}
\newcommand{\Th}{\Theta}
\newcommand{\om}{\omega}
\newcommand{\be}{\begin{equation}}
\newcommand{\ee}{\end{equation}}
\newcommand{\beq}{\begin{eqnarray}}
\newcommand{\eeq}{\end{eqnarray}}
\newcommand{\lc}{\varepsilon}
\newcommand{\al}{\alpha}
\newcommand{\Ga}{\Gamma}
\newcommand{\si}{\sigma}
\newcommand{\fhi}{\varphi}

\begin{document}

\title{EXTENDING THE BARNES-RIVERS OPERATORS TO D=3 TOPOLOGICAL GRAVITY.}

\vspace{2cm}

\author{ C. Pinheiro, \\
Universidade Federal do Esp\'\i rito Santo, \\
Instituto de F\'\i sica e Qu\'\i mica, \\
Av. Fernando Ferrari, s/n., Campus Goiabeiras, \\
29069 Vit\'oria, E.S., Brazil \\
and \\
Universidade Federal do Rio de Janeiro, \\
Instituto de F\'\i sica, \\
21944 Rio de Janeiro, R.J., Brazil \\
 and \\ \and
G. O. Pires,
\\Centro Brasileiro de Pesquisas F\'\i sicas, \\
Departamento de Campos e Part\'\i culas, \\
Rua Dr. Xavier Sigaud, 150, Urca,\\
22290 Rio de Janeiro, R.J.,Brazil.  }

\footnotetext[1]{BitNet address   : Gentil@BRLNCC.BITNET }
\footnotetext[2]{InterNet address : Gentil@CBPFSU1.CAT.CBPF.BR }

\date{October, 1992.}

\maketitle

\vspace{2cm}

\begin{abstract}

{\it The spin-projector operators for symmetric rank-2 tensors are  reassessed
in connection with the issue of topologically massive gravity. The original
proposal by Barnes and Rivers is generalised to account for D-dimensional
Einstein gravity and 3-dimensional Chern-Simons massive gravitation.}

\end{abstract}

\pagebreak

\vspace{1cm}

The possibility of building up a quantum-mechanically consistent gauge
theory for the gravitational field seems to be actually realised in
3-dimensional space-time.
The early work by Deser, Jackiw and Templeton \cite{DJT} brings about the issue
of a massive dynamical theory for gravitation in 3D. Ever since,
topologically massive gravity has been fairly-well investigated in a
series of very interesting papers, till very recently it has been shown
that it is not only renormalisable \cite{DY,Gainesville} but even more :
massive
3D-gravity is a finite quantum field theory \cite{PS}.

The purpose of this letter is to reassess the set of Barnes-Rivers spin
operators \cite{Rivers,Barnes}  in the framework of topologically massive
gravity.
These have been shown to be very relevant in the description of
4D-quantum gravity \cite{N,AT}. We shall in this letter propose a set of
operators that extend the original Barnes-Rivers projectors to
include D-dimensional massless and massive gravity as well as
3D-gravity  with topological mass.
The graviton propagators shall be written down and the tree-level
unitarity shall be discussed in terms of the residues of the propagators
at their  poles.

\vspace{.3cm}

The Barnes-Rivers spin-projectors, as introduced in  \cite{Rivers,Barnes},
form a complete set of spin-projector operators in the space of rank-2 tensors.
For the symmetric case, they read as below :

\vspace{1.0cm}

\[
\begin{array}{lll}
P^{(2)}_{\mu \nu, \ka \la} & \equiv &  \frac{1}{2} ( \Theta_{\mu \ka}
\Theta_{\nu \la} + \Theta_{\mu \la} \Theta_{\nu \ka} ) - \frac{1}{3}
\Theta_{\mu \nu} \Theta_{\ka \la} ,
 \\ & &  \\
P^{(1)}_{\mu \nu, \ka \la} & \equiv &  \frac{1}{2} ( \Th_{\mu \ka} \om_{\nu
\la}
+ \Th_{\mu \la} \om_{\nu \ka} + \Th_{\nu \ka} \om_{\mu \la} +
\Th_{\nu \la} \om_{\mu \ka} ),
 \\ & & \\
P^{(0)}_{s \:\:\mu \nu, \ka \la} & \equiv &  \frac{1}{3} \Th_{\mu \nu} \Th_{\ka
\la},
 \\ & & \\
P^{(0)}_{w \:\:\mu \nu, \ka \la} & \equiv &  \om_{\mu \nu} \om_{\ka \la},
 \\ & & \\
P^{(0)}_{sw \:\:\mu \nu, \ka \la} & \equiv &  \frac{1}{\sqrt{3}} \Th_{\mu \nu}
\om_{\ka \la},
\\ & & \\
P^{(0)}_{ws \:\:\mu \nu, \ka \la} & \equiv &  \frac{1}{\sqrt{3}} \om_{\mu \nu}
\Th_{\ka \la},
\end{array}
\:\:\:\:\:\:\:\:\:\:\:\:\:\:\:\:\:\:\:\:\:\:\:\:\:\:\:\:\:\:\:\:\:\:\:\:\:\:\:
\begin{array}{r}
(1.a)\\  \\
(1.b)\\  \\
(1.c)\\  \\
(1.d)\\  \\
(1.e)\\  \\
(1.f)
\end{array}
\]

\vspace{1cm}

\noindent where $ \Th_{\mu \nu} $ and $ \om_{\mu \nu} $ are the usual
 transverse and longitudinal projectors on the space of vectors.
The operators in (1.a) and (1.b) are respectively the \mbox{spin-2} and -1
projectors. The remaining ones project out \mbox{spin-0} components of
rank-2 symmetric tensors.

Let us now consider the Einstein-Hilbert action for gravitation and derive
its propagator by means of the algebra of the Barnes-Rivers operators, taken
now in a D-dimensional space-time :

\setcounter{equation}{1}

\be
{\cal L}_{HE} = \frac{1}{2 \ka^{2}} \sqrt{-g} {\cal R}. \;
\ee

\noindent Adopting the viewpoint of expanding the metric field around the
 flat-space geo\-metry, \footnote[3]{diag. $ \eta^{\mu \nu} \equiv (+\, ; - \,
, \cdots
 \, , \, -). $ }

\be
g^{\mu \nu}(x)=\eta^{\mu \nu} - \ka h^{\mu \nu}(x),
\ee

\noindent where $ h^{\mu \nu} $ is the field variable defining the expansion,
and taking into account only the free sector of the expansion, one gets the
following free Lagrangean for the  $ h^{\mu \nu} $-field :

\be
{\cal L}^{free}_{HE}  =  \frac{1}{4} \partial_{\la} h_{\mu \nu}
\partial^{\la} h^{\mu \nu} - \frac{1}{4} \partial_{\la} h^{\mu}_{\;\;\mu}
\partial^{\la} h^{\nu}_{\;\;\nu}+ \frac{1}{2} \partial_{\la} h^{\la}_{\;\;\mu}
\partial^{\mu} h^{\nu}_{\;\;\nu}- \frac{1}{2} \partial_{\la} h^{\la}_{\;\;\mu}
 \partial_{\nu} h^{\nu \mu}.
\ee

To give meaning to the integration measure in the generating functional of
Green's functions, it is necessary to fix the gauge invariance

\be
\delta h_{\mu \nu}(x) = \partial_{\mu} \zeta_{\nu}(x)  + \partial_{\nu}
\zeta_{\mu}(x),
\ee

\noindent by introducing the De Donder  gauge-fixing term :

\[
\begin{array}{rrrlr}
\:\:\:\:\:\:\:\:\:\:\:\:\:\:\:\:\:\:\:\:\:
& {\cal L}_{g.f.} & = & \frac{1}{2 \alpha} F_{\mu} F^{\mu} \; ,
&
\:\:\:\:\:\:\:\:\:\:\:\:\:\:\:\:\:\:\:\:\:\:\:\:\:\:\:\:\:\:\:\:\:\:\:\:\:\:\:\:\:\:\:   (6.a)
\\ \mbox{where}
\:\:\:\:\:\:\:\:\:\:\:\:\:\:\:\:\:\:\:\:\:\:\:\:\:\:\:\:\:\:  & & & &  \\
\:\:\:\:\:\:\:\:\:\:\:\:\:\:\:\:\:\:\:\:\:
& F_{\mu} [h_{\rho \si}] & = & \partial_{\la} ( h^{\la}_{\;\;\mu} - \frac{1}{2}
\delta^{\la}_{\mu} h^{\nu}_{\;\;\nu} ).
&
\:\:\:\:\:\:\:\:\:\:\:\:\:\:\:\:\:\:\:\:\:\:\:\:\:\:\:\:\:\:\:\:\:\:\:\:\:\:\:\:\:   (6.b)
\end{array}
\]

The Hilbert-Einstein Lagrangean with gauge-fixing term can be rewritten in
terms of the operators (1.a)-(1.f) according to :

\setcounter{equation}{6}

\be
{\cal L}^{(2)} = \frac{1}{2} h^{\mu \nu} {\cal O}_{\mu \nu, \ka \la} h^{\ka
\la},
\ee

\noindent where

\beq
{\cal O}_{\mu \nu, \ka \la} & = & \Box \left( - \frac{1}{2} P^{(2)} -
\frac{1}{2 \al} P^{(1)}_{m} + \frac{(4 \al - 3)}{4 \al} P^{(0)}_{s} +
\right.\nonumber  \\  & & \left.
 - \frac{1}{4 \al} P^{(0)}_{w} + \frac{\sqrt{3}}{4 \al} P^{(0)}_{sw} +
\frac{\sqrt{3}}{4 \al} P^{(0)}_{ws} \right)_{\mu \nu , \ka \la}.
\eeq

The associated propagator is obtained from  the generating-functional

\be
{\cal W}[\tau_{\rho \sigma}] = - \frac{1}{2} \int d^{D}\!x \; d^{D}\!y \;
\tau^{\mu \nu} {\cal O}^{-1}_{\mu \nu , \ka \la} \tau^{\ka \la},
\ee

\noindent so that :

\be
<\;T \; [h_{\mu \nu} (x) \; h_{\ka \la} (y)\;]\; > = i {\cal O}^{-1}_{\mu \nu ,
\ka \la} \delta^{D}(x-y) . \;
\ee

\noindent  So, using the rank-2 identity in the space of symmetric
\mbox{rank-2} tensors, one gets :

\beq
<\;T\;[ h_{\mu \nu} (x) \; h_{\ka \la} (y)\;] \;>  & = & \frac{i}{\Box} \left
\{
-2 P^{(2)} -2 \alpha P^{(1)}_{m} - 2 \frac{(D-5)}{(D-2)} P^{(0)}_{s} +
\right.\nonumber \\  & & \left. - 2 \frac{(2D\alpha -4\alpha -D+1)}{(D-2)}
P^{(0)}_{w} + 2 \frac{\sqrt{3}}{(D-2)} P^{(0)}_{sw} + \right.\nonumber \\  & &
\left.
+ 2 \frac{\sqrt{3}}{(D-2)} P^{(0)}_{ws} \right \}_{\mu \nu , \ka \la}
\delta^{D}(x-y), \;
\eeq

\noindent or, in momentum space :

\beq
<\; T \;[ h_{\mu \nu} (-k) \; h_{\ka \la} (k)\;] \; > & = & \frac{i}{k^{2}}
\left \{
\eta_{\mu \ka} \eta_{\nu \la} + \eta_{\mu \la} \eta_{\nu \ka} -
\frac{2}{(D-2)} \eta_{\mu \nu} \eta_{\ka \la} +
\right.\nonumber \\  & & \left.
-(1-\al) \left[ \eta_{\mu \ka} \om_{\nu \la} + \eta_{\nu \ka}
\om_{\mu \la} + \eta_{\mu \la} \om_{\nu \ka} \right] \right \}, \;
\eeq

\noindent where the projectors have been replaced by eqs. (1.a)-(1.f),
and the gauge-fixing parameter has been kept arbitrary.

Adding to the Hilbert-Einstein action a Proca-like mass term yields the
fol\-low\-ing expression for the graviton propagator :

\beq
<\; T \;[ h_{\mu \nu} (-k) \; h_{\ka \la} (k)\;] \; > & = &
\frac{i}{(k^{2}-m^{2})} \left \{ \; \left( \eta_{\mu \ka} \eta_{\nu \la} +
\eta_{\mu \la} \eta_{\nu \ka} + \frac{2}{(D-1)} \eta_{\mu \nu} \eta_{\ka \la}
\right) \right. + \nonumber  \\
 & & \!\!\!\!\!\!\!\!\!\!\!\!\!\!\!\!\!\!\!\!\!\!\!\!\!\!\!\!
 + \; 2 \; i \left[ \frac{(k^{2}-m^{2}) \, [ \, (2m^{2}-1)(D-1)+1 \,] -m^{4}D}{
m^{4} ( D-1 )} \right]  \om_{\mu \nu} \om_{\ka \la} + \nonumber \\
&  & \!\!\!\!\!\!\!\!\!\!\!\!\!\!\!\!\!\!\!\!\!\!\!\!\!\!\!\!\!
- \; \frac{ \, i \; k^{2}}{m^{2}} \left[ \eta_{\mu \ka} \om_{\nu \la} +
\eta_{\mu \la} \om_{\nu \ka} + \eta_{\nu \la} \om_{\mu \ka}  +
\eta_{\nu \ka} \om_{\mu \la} + \right. \nonumber \\
& & \!\!\!\!\!\!\!\!\!\!\!\!\!\!\!\!\!\!\!\!\!\!\!\!\!\!\!\!
\left. \left. + \; \frac{2}{(D-1)} \left( \eta_{\mu \nu} \om_{\ka \la} +
  \eta_{\ka \la} \om_{\mu \nu} \right)  \right] \; \right \} \, .
\eeq

\vspace{.3cm}

An extension of the Barnes-Rivers operators can be proposed in order to
ac\-count for D=3 topologically massive gravity \cite{DJT}. It can be
shown that one needs to add two operators to the Table I,

\[
S_{1 \; \mu \nu , \ka \la}  \equiv  \frac{(- \Box)}{4} \left \{
\lc_{\mu \al \la} \partial_{\ka} \om^{\al}_{\;\;\nu} +
\lc_{\mu \al \ka} \partial_{\la} \om^{\al}_{\;\;\nu} +
\lc_{\nu \al \la} \partial_{\ka} \om^{\al}_{\;\;\mu} +
\lc_{\nu \al \ka} \partial_{\la} \om^{\al}_{\;\;\mu}
\right \} \; \mbox{\hspace{1.2cm} (14.a)}
\]

\noindent and

\[
S_{2 \; \mu \nu , \ka \la}  \equiv  \frac{\Box}{4} \left \{
\lc_{\mu \al \la} \eta_{\ka \nu} +
\lc_{\mu \al \ka} \eta_{\la \nu} +
\lc_{\nu \al \la} \eta_{\ka \mu} +
\lc_{\nu \al \ka} \eta_{\la \mu}
\right \} \partial^{\al} , \; \mbox{\hspace{3.1cm} (14.b)}
\]

\setcounter{equation}{14}

\noindent which can be found by analysing  the bilinear part
 stemming from the gravitational Chern-Simons term :

\be
{\cal L}_{CS} =  \frac{1}{\mu} \lc^{\la \mu \nu}
\Ga^{\rho}_{\;\;\la \si} ( \partial_{\mu} \Ga_{\rho \;\;\nu}^{\;\;\si} +
\frac{2}{3} \Ga_{\mu \;\;\fhi}^{\;\;\si} \Ga^{\;\;\fhi}_{\nu \;\;\rho} ).
\;
\ee

Fixing the gauge as in (6), the bilinear term coming from the
Hilbert-Einstein and Chern-Simons actions looks as follows :

\beq
{\cal L}^{(2)} & = & \frac{1}{2} h^{\mu \nu} \left \{ \Box \left [
\frac{1}{2} P^{(2)} + \frac{1}{2 \al} P^{(1)}_{m} -
\frac{(4 \al -3)}{4 \al} P^{(0)}_{s} + \right. \right.\nonumber \\
 &  & \left. + \frac{1}{4 \al} P^{(0)}_{w}
-\frac{\sqrt{3}}{4\al} P^{(0)}_{sw}
-\frac{\sqrt{3}}{4\al} P^{(0)}_{ws} \right ] +\nonumber \\
& & \left. + 4 (\frac{\ka^{2}}{\mu}) [ S_{1} + S_{2} ] \right \}_{\mu \nu ,
\ka \la} h^{\ka \la}. \;
\eeq

Again, the associated propagator can be read off with the help of the operator
algebra displayed in \mbox{Table I} :

\beq
<\;T\;[ h_{\mu \nu} (x) \; h_{\ka \la} (y)\;]\; > & = & \frac{i}{\Box} \left \{
\frac{2(\frac{\mu}{\ka^{2}})^{2} }{[(\frac{\mu}{\ka^{2}})^{2}+64\Box]}
P^{(2)} + 2 \al P^{(1)}_{m} \right. + \nonumber  \\
- \frac{4 [ (\frac{\mu}{\ka^{2}})^{2} + 48 \Box ] }{
[(\frac{\mu}{\ka^{2}})^{2}+64\Box]}P^{(0)}_{s} & + & 4 (\al -1 ) P^{(0)}_{w}
-      2 \sqrt{3} P^{(0)}_{sw} + \nonumber  \\
-     2 \sqrt{3} P^{(0)}_{ws}
& - & \left. \frac{16 (\frac{\mu}{\ka^{2}}) }{\Box [(\frac{\mu}{\ka^{2}})^{2}
 + 64 \Box ]} [ S_{1} + S_{2} ] \right \}_{\mu \nu,\ka \la}\delta^{3}(x-y).
 \;
\eeq

\noindent By choosing $ \al = 1 $ ( Feynman gauge ), we can write in momentum
 space :

\beq
<\;T\;[ h_{\mu \nu} (-k) \; h_{\ka \la}(k)\;] \;> & = & \frac{-i}{k^{2} \,
 [64 k^{2}-(\frac{\mu}{\ka^{2}})^{2}]  } \left\{
 4 \,i \, (\frac{\mu}{\ka^{2}}) k^{\al} \,
\left [ \lc_{\mu \al \la} \Th_{\ka \nu} + \lc_{\mu \al \ka} \Th_{\la \nu} +
\right. \right.  \nonumber \\  & + & \left.
 \lc_{\nu \al \la}\Th_{\ka \mu} + \lc_{\nu \al \ka}\Th_{\la \mu} \right ] +
\nonumber   \\  & - &
  (\frac{\mu}{\ka^{2}})^{2}
\left [ \eta_{\mu \ka} \eta_{\nu \la} + \eta_{\mu \la} \eta_{\nu \ka} -
2 \eta_{\mu \nu} \eta_{\ka \la} \right ] +
\nonumber \\ & - &
 64 k^{2} \left[ \eta_{\mu \ka} \om_{\nu \la} + \eta_{\mu \la} \om_{\nu \ka} +
\eta_{\nu \ka} \om_{\mu \la} +
\right.  \nonumber \\ & + & \left. \left.
 \eta_{\nu \la} \om_{\mu \ka} + \Th_{\mu \nu} \Th_{\ka \la} -
2 \eta_{\mu \nu} \om_{\ka \la} - 2 \eta_{\ka \la} \om_{\mu \nu} \right ] \,
\right \} .
\eeq

As it can be seen, the Hilbert-Einstein action in D=4 leads to a massless
dynamical pole in the $h_{\mu \nu}$-propagators, whereas the
Einstein-Chern-Simons D=3-action yields 2 poles : a massless non-dynamical
excitation along with a non-tachyonic massive dynamical mode,

\be
k^{2} = (\frac{\mu}{8 \ka^{2}})^{2} > 0  ,
\ee

\noindent as already known from \cite{DJT}.

\vspace{.3cm}

Coupling the propagator to external currents, $ \tau^{\mu \nu} $,
compatible with the  sym\-metries of the theory, and then
taking the imaginary part of the residues of the amplitude at the poles,
one can probe the necessary condition for unitarity at tree-level and count
degrees of freedon described by the field. The current-current transition
amplitude in  momentum space is written as :

\be
{\cal A} \equiv \tau^{\ast \: \mu \nu}(k) <\;T\;[ h_{\mu \nu} (-k) \; h_{\ka
\la} (k)\;]\; >
\tau^{\ka \la}(k) \; , \:
\ee

\noindent where only the spin-projectors $ P^{(2)}, P^{(0)}_{s} $ and $ S_{2} $
shall contribute due to the trans\-ver\-sality of $ \tau^{\mu \nu}(k) $.
Now, defining the following set of independent vectors in momentum space :

\be
\left\{  \begin{array}{lll}
k^{\mu} & \equiv & ( k^{0} ; \vec{k} ) \\
 \tilde{k}^{\mu} & \equiv &  ( k^{0} ; -\, \vec{k} ) \:  \\
 \lc^{\mu}_{i} & \equiv & ( 0 ; \vec{\lc}_{i} ) \: , \: i=1 \ldots D-2,
\end{array} \right.
\ee

\noindent we can write the symmetric current tensor $ \tau^{\mu \nu}(k) $
as

\beq
\tau_{\mu \nu}(k) & = & a(k) k_{\mu} k_{\nu} +
 b(k) k_{(\mu} \tilde{k}_{\nu)} + c_{i}(k) k_{(\mu} \lc^{i}_{\nu)} +
\nonumber\\
& & + \; d(k) \tilde{k}_{\mu} \tilde{k}_{\nu} +
e_{i}(k)\tilde{k}_{(\mu}\lc^{i}_{\nu)}
 + f_{ij}(k) \lc^{i}_{(\mu} \lc^{j}_{\nu)} \; ,
\eeq

\noindent and then extract some relations involving the above coeficients
when imposing its conservation for on-shell momenta $ k^{\mu} $.

\vspace{.3cm}

So, for the Einstein theory in D dimensions, the amplitude $ {\cal A} $
\mbox{reads :}

\be
{\cal A} = \frac{(-i)}{k^{2}} \tau_{\mu \nu}^{\ast}(k) \left \{ -2 P^{(2)}(k)
+ \frac{2 (5-D)}{(D-2)} P^{(0)}_{s}(k) \right \}^{\mu \nu, \ka \la}
\tau_{\ka \la}(k) \: ;
\ee

\noindent then, at the pole $ k^{2}=0 $,

\be
Im \; Res \; {\cal A}
 = \left[ 2 \vert \tau_{\mu \nu} \vert^{2} - \frac{2}{3}
\left( 1 + \frac{5-D}{D-2} \right) \vert \tau^{\mu}_{\:\:\mu} \vert^{2}
\right].
\:
\ee

\noindent Manipulating with $ \tau_{\mu \nu}(k) $ as expanded above, one gets :

\be
Im \; Res \; {\cal A} = 2 \: \left[ \vert f_{ij} \vert^{2} - \frac{1}{3}
\left( 1 + \frac{5-D}{D-2} \right) \vert f_{ii} \vert^{2} \right]. \:
\ee

\noindent For D=4 dimensions,

\be
Im \; Res \; {\cal A} = 2 \: \left[ \frac{1}{2} \vert f_{11}-f_{22} \vert^{2}
 + 2 \vert f_{12} \vert^{2} \right] \:\: > \:\: 0. \:
\ee

\noindent Upon solving the eingenvalue problem of the M-matrix of (26) :

\be
\frac{1}{2} \: Im \: Res \; {\cal A} =
\left( \begin{array}{lll}
          f_{11}^{\ast} & f_{22}^{\ast} & f_{12}^{\ast}
       \end{array}
\right) \:
\underbrace{ \left( \begin{array}{ccc}
          \frac{1}{2}  & \frac{-1}{2} & 0 \\
          \frac{-1}{2} & \frac{1}{2}  & 0 \\
                0      &      0       & 2
       \end{array}
\right) }_{M} \:
\left( \begin{array}{c}
          f_{11} \\ f_{22} \\ f_{12}
       \end{array}
\right), \:\:
\ee

\noindent one gets two non-vanishing eingenvalues that describe the two
\mbox{on-shell} degrees of freedom of the massless graviton.

\pagebreak

\noindent For D=3 dimensions,

\be
Im \: Res \: {\cal A} = 2 \left( \vert f_{ij} \vert^{2} -
 \vert f_{ii} \vert^{2} \right) = 0,\: \:\:( i=j=1),
\ee

\noindent confirming, as it is known, that the \mbox{Einstein} theory is
non-dy\-na\-mi\-cal in 3 di\-men\-sions.

\vspace{.3cm}

For the Einstein-Chern-Simons theory,

\beq
{\cal A} & = & \frac{i}{ k^{2} [ 64k^{2}-(\frac{\mu}{\ka^{2}})^{2} ]} \:
\tau_{\mu \nu}^{\ast}(k) \; \left \{ 2 \; (\frac{\mu}{\ka^{2}})^{2} \,
P^{(2)}(k) +
 \right.\nonumber \\
& - &  \left. [ 4 (\frac{\mu}{\ka^{2}})^{2} - 192 k^{2} ] \, P^{(0)}_{s}(k) +
 \frac{16 (\frac{\mu}{\ka^{2}}) }{k^{2}} \,
 S_{2}(k) \right \}^{\mu \nu, \ka \la}
\; \tau_{\ka \la}(k).
\eeq

\noindent At the pole $ k^{2}=0 $,

\begin{eqnarray}
Im \; Res \; {\cal A} & = &  \lim_{k^{2} = 0} \;
\frac{1}{ [ 64 k^{2}-(\frac{\mu}{\ka^{2}})^{2} ] } \left \{
2 (\frac{\mu}{\ka^{2}})^{2} \,
\left[ \vert \tau_{\ka \la} \vert^{2} - \frac{1}{3} \vert \tau^{\mu}_{\:\:\mu}
\vert^{2} \right] \right. + \nonumber \\
& - & \left. \frac{ 4[(\frac{\mu}{\ka^{2}})^{2}-48 k^{2}]}{3}
\, \vert \tau^{\mu}_{\:\:\mu} \vert^{2} \, +
 16 (\frac{\mu}{\ka^{2}}) \, k^{\al} \,
\lc_{\mu \al \la} \tau^{\ast \, \mu}_{\:\:\:\:\:\ka} \: \tau^{\ka \la}
\right \} \nonumber \\
& = & \lim_{k^{2}=0} \left \{ \frac{ 64 k^{2} \vert f \vert^{2} }{[ 64
k^{2}-(\frac{\mu}{\ka^{2}})^{2} ]}
 \right \} = 0;
\end{eqnarray}

\noindent which is therefore shown to be non-propagating.

\vspace{.3cm}

\noindent At the pole $ k^{2}= (\frac{\mu}{ 8 \ka^{2}})^{2} $,

\begin{eqnarray}
Im \; Res \; {\cal A} & = &  \lim_{k^{2} = (\frac{\mu}{ 8 \ka^{2}})^{2} } \,
\frac{1}{k^{2}} \left \{
\, 2 (\frac{\mu}{\ka^{2}})^{2}
\, \left[ \vert \tau_{\ka \la} \vert^{2} -
\frac{1}{3} \vert \tau^{\mu}_{\:\: \mu} \vert^{2} \right] \right. + \nonumber
\\
& - & \left. \frac{4 [ (\frac{\mu}{\ka^{2}})^{2}-48 k^{2}] }{ 3 }
\, \vert \tau^{\mu}_{\:\: \mu} \vert^{2} \, +
\,  16 \, (\frac{\mu}{ \ka^{2}}) \, k^{\al} \, \lc_{\mu \al \nu}
\tau^{\ast \, \mu}_{\:\:\:\:\: \ka} \; \tau^{\ka \la}
\right \} \nonumber \\
& = & 64 \vert f \vert^{2} \: > \: 0;
\end{eqnarray}

\noindent giving one degree of freedom. Here, attention must be paid to the
sign of the Hilbert-Einstein Lagrangean in D=3 : a minus sign has to be  chosen
in
(16) in order to guarantee a ghost-free massive propagator in three dimensions,
 although, with our choice of metric, the opposite sign is the one needed
 to ensure that the massless graviton is not a ghost.

\vspace{.3cm}

To conclude, we have set the spin-projector  operators to deal with
D-di\-men\-sional Einstein's gravity and
D=3-topologically massive gravitation. Their multiplicative table
has been used in the derivation of the graviton propagators in a general
gauge.

Having in mind the coupling of a Maxwell-\-Chern-\-Si\-mons gau\-ge fi\-eld to
\mbox{Einstein}-\mbox{Chern}-\mbox{Simons} gravity, the propagator (17) will be
employed to
explicitly calculate one-loop corrections to the coupled gauge-gravity
system. These results shall be presented and discussed in a further work
\cite{next}.

\vspace{1cm}

We are indebted to Dr. J. A. Helay\"el-Neto for patient discussions and
careful guidance. We are also grateful to O. M. Del Cima
for a helpful discussion. We express our gratitude to the members of the
Theoretical Physics Group of the Universidade Cat\'olica de Petr\'opolis
for kind hospitality. M. A. Andrade is acknowledged for the kind help
in typesetting with \LaTeX. The authors are grateful to the CNPq. and CAPES
for the invaluable financial support.

\pagebreak

\vspace{2cm}

\underbar{{\bf APPENDIX I }}

\vspace{.5cm}

Multiplicative Table for the Barnes-Rivers spin-projector
operators in D di\-men\-sions :

\vspace{.7cm}

\begin{eqnarray*}
P^{(2)} P^{(2)}& = & P^{(2)} + \frac{(D-4)}{3} P^{(0)}_{s},\\
P^{(1)}_{m} P^{(1)}_{m} & = & P^{(1)}_{m},\\
P^{(2)} P^{(0)}_{s} & = & \frac{(4-D)}{3} P^{(0)}_{s},\\
P^{(0)}_{s} P^{(2)} & = & \frac{(4-D)}{3} P^{(0)}_{s},\\
P^{(2)} P^{(0)}_{sw} & = & \frac{(4-D)}{3} P^{(0)}_{sw},\\
P^{(0)}_{ws} P^{(2)} & = & \frac{(4-D)}{3} P^{(0)}_{ws},\\
P^{(0)}_{s} P^{(0)}_{s} & = & \frac{(D-1)}{3} P^{(0)}_{s},\\
P^{(0)}_{w} P^{(0)}_{w} & = &  P^{(0)}_{w},\\
P^{(0)}_{s} P^{(0)}_{sw} & = & \frac{(D-1)}{3} P^{(0)}_{sw},\\
P^{(0)}_{ws} P^{(0)}_{s} & = & \frac{(D-1)}{3} P^{(0)}_{ws},\\
P^{(0)}_{sw} P^{(0)}_{w} & = &  P^{(0)}_{sw},\\
P^{(0)}_{w} P^{(0)}_{ws} & = &  P^{(0)}_{ws},\\
P^{(0)}_{sw} P^{(0)}_{ws} & = &  P^{(0)}_{s},\\
P^{(0)}_{ws} P^{(0)}_{sw} & = & \frac{(D-1)}{3} P^{(0)}_{w}.
\end{eqnarray*}

\vspace{1.0cm}

Extension to the case of 3D massive gravity :

\vspace{.7cm}

\begin{eqnarray*}
S_{2} S_{2} & = & \Box^{3} \left\{ \frac{1}{2} P^{(0)}_{s} -
\frac{1}{4} P^{(1)}_{m} -  P^{(2)} \right\},\\
S_{2} S_{1} & = & \frac{\Box^{3}}{4} P^{(1)}_{m},\\
S_{1} S_{2} & = & \frac{\Box^{3}}{4} P^{(1)}_{m},\\
S_{1} S_{1} & = & \frac{(-\Box^{3})}{4} P^{(1)}_{m},\\
S_{1} P^{(1)}_{m} & = & S_{1},\\
P^{(1)}_{m} S_{1} & = & S_{1},\\
S_{2} P^{(2)} & = & S_{2} + S_{1},\\
P^{(2)} S_{2} & = & S_{2} + S_{1},\\
S_{2} P^{(1)}_{m} & = & -\; S_{1},\\
P^{(1)}_{m} S_{2} & = & -\;S_{1}.
\end{eqnarray*}

\vspace{1cm}

Tensorial identity :

\vspace{.7cm}

\[
\left \{ P^{(2)} + P^{(1)} + P^{(0)}_{s} + P^{(0)}_{w} \right \}_{\mu \nu, \ka
\la}
= \frac{1}{2} \left( \eta_{\mu \ka} \eta_{\nu \la} +
\eta_{\mu \la} \eta_{\nu \ka} \right).
\]

\end{document}